\documentclass[%
aip,
jcp,
amsfonts,
amsmath,
amssymb,
reprint,
]{revtex4-1}
\usepackage[colorlinks,linkcolor=blue,urlcolor=blue,citecolor=blue]{hyperref}
\usepackage{bm}
\usepackage{graphicx}
\usepackage{subfig}

\begin{document}

\title{Refining Potential Energy Surface through Dynamical Properties via Differentiable Molecular Simulation}
\author{Bin Han}
\affiliation{Institute of Materials Research, Tsinghua Shenzhen International Graduate
 School (TSIGS), Shenzhen, 518055, People’s Republic of China}
\author{Kuang Yu}
\email[Email: ]{yu.kuang@sz.tsinghua.edu.cn}
\affiliation{Institute of Materials Research, Tsinghua Shenzhen International Graduate
 School (TSIGS), Shenzhen, 518055, People’s Republic of China}

\begin{abstract}

Recently, machine learning potentials (MLP) largely enhances the reliability of molecular dynamics, but its accuracy is limited by the underlying \textit{ab initio} methods. A viable approach to overcome this limitation is to refine the potential by learning from experimental data, which now can be done efficiently using modern automatic differentiation technique. However, potential refinement is mostly performed using thermodynamic properties, leaving the most accessible and informative dynamical data (like spectroscopy) unexploited. In this work, through a comprehensive application of adjoint and gradient truncation methods, we show that both memory and gradient explosion issues can be circumvented in many situations, so the dynamical property differentiation is well-behaved. Consequently, both transport coefficients and spectroscopic data can be used to improve the density functional theory based MLP towards higher accuracy. Essentially, this work contributes to the solution of the inverse problem of spectroscopy by extracting microscopic interactions from vibrational spectroscopic data. 

\end{abstract}

\maketitle

\section*{Introduction}
In chemistry and materials science, learning from experimental data is crucial for expediting  the design of new materials. Experimental observables, such as density, diffusion coefficient, and dielectric constant, provide valuable insights into atomic interactions within materials. Among these properties, dynamical data plays a pivotal role: transport properties (e.g., diffusion coefficients, electrical conductivity, and thermal conductivity etc.) are important prediction targets in many application scenarios. While vibrational spectroscopic data contains comprehensive microscopic information, making it an ideal probe to unravel the structures and the dynamics of the system at atomic level. However, the analysis and interpretation of dynamical data, especially spectra, has been quite subjective and prone to human error. Therefore, establishing a direct and rigorous connection between spectra and microscopic dynamics has always been a great challenge.

For long time, molecular dynamics (MD) simulation has been a powerful tool to predict, analyze, and interpret dynamical properties. For example, both equilibrium and nonequilibrium MD techniques can be applied to predict transport properties such as diffusion coefficients\cite{wheelerMolecularDynamicsSimulations2004a,wheelerMolecularDynamicsSimulations2004}. MD simulations  can reach excellent agreement with experimental measurements across a wide range of temperatures and pressures, given that a correct potential model is used\cite{tsimpanogiannisSelfdiffusionCoefficientBulk2019}. Moreover, since the 1990s\cite{guillotMolecularDynamicsStudy1991}, MD simulation has served as a standard technique to study infrared (IR) spectrum, unveiling valuable physical insights, such as the connection between the hydrogen-bond stretch peak at 200 cm$^{-1}$ with the intermolecular charge transfer effects in liquid water\cite{sharmaIntermolecularDynamicalCharge2005,hanIncorporatingPolarizationCharge2023}. However, similar issue exists here that the dynamical simulation results are sensitive to the underlying potential energy surface (PES). Consequently, different PES give significantly different predictions, making the reliability of the corresponding interpretations under debate. 

Conventionally, classical potentials used in MD simulations rely on physically motivated functional forms to describe inter-atomic interactions. The accuracy of classical potentials thus are largely limited by the predetermined functional form. In contrast, in recent years, machine learning (ML) methods are quickly gaining popularity due to their flexibility and consequently, excellent fitting capability in high-dimensional space.  Models such as BPNN\cite{behlerGeneralizedNeuralNetworkRepresentation2007,behlerConstructingHighDimensional2015}, DeepPMD\cite{zhangDeepPotentialMolecular2018}, EANN\cite{zhangEmbeddedAtomNeural2019}, NequIP\cite{batznerEquivariantGraphNeural2022} etc have been developed and used in various circumstances. Besides fitting PES, ML methods also demonstrate extraordinary capability in predicting tensorial properties such as dipole moment and polarizability\cite{zhangDeepNeuralNetwork2020,zhangUniversalMachineLearning2023,chenAcceleratingMolecularVibrational2024,xuTensorialPropertiesNeuroevolution2024a,chengDevelopingDifferentiableLongRange2024}, which are used to generate highly accurate IR and Raman spectra. However, to date, most ML potentials are fitted to \textit{ab initio} energies and forces, the accuracy of which is inevitably limited by the underlying \textit{ab initio} method. Only a few works are based on the chemically accurate "golden standard" coupled cluster theory\cite{babinDevelopmentFirstPrinciples2013a,smithApproachingCoupledCluster2019,yangTransferrableRangeseparatedForce2022}, while most ML potentials are still fitted to much cheaper density functional theory (DFT) calculations due to the limitation of training cost. Worsening the situation, most condensed phase potentials rely on condensed phase training data, for which massive accurate correlated wavefunction calculations are prohibited. The cheap pure functionals typically used in extended system calculations are usually not accurate enough to describe molecular interactions. 

In this work, instead of predicting dynamical properties using bottom-up ML potential, we focus on the inverse problem: how to refine the underlying potentials using dynamical data, especially spectroscopic data. Compared to the prediction task, the inverse problem is actually more in alignment with the fundamental question of spectroscopy: that is, how to infer detailed microscopic information of the system from its spectroscopy? In fact, such "top-down" strategy\cite{frohlkingEmpiricalForceFields2020} is quite common in classical force field (FF) development, but mostly using thermodynamic data such as densities and evaporation enthalpies and leaving the most accessible and informative spectroscopic data unexploited. As stated above, the difficulty lies in the fact that the quantitative mapping between PES and spectroscopy is hard to establish. In most existing works, people have to guess the form of PES, and manually adjust parameters to match the spectroscopy, which is highly non-scalable, especially in the era of ML potentials. However, recently, the emergence of automatic differentiation (AD) technique enables extremely efficient and automatic parameter optimization, making this task possible.  

Utilizing the AD  technique, there are several successful examples of top-down fitting implementations. Unlike a direct end-to-end matching between structure and property\cite{luDeepLearningAssistedSpectrum2024}, fitting via differentiable MD simulation can yield physically meaningful potential, which is transferable to the prediction of other properties. Using experimental data to correct bottom-up fitting results has been proven effective in many works\cite{matinMachineLearningPotentials2024,rockenAccurateMachineLearning2024,gongBAMBOOPredictiveTransferable2024}, supporting  a new PES development strategy: pre-train the model using a cheap \textit{ab initio} method, then fine-tune it with a small amount of experimental data. Such strategy can significantly lower the training cost and eases the construction of ML potentials. To date, several infrastructures for differentiable MD simulation have been implemented, including JAX-MD\cite{schoenholzJAXFrameworkDifferentiable2020}, TorchMD\cite{doerrTorchMDDeepLearning2021}, SPONGE\cite{huangSPONGEGPUAccelerated2022} and DMFF\cite{wangDMFFOpenSourceAutomatic2023}. Building upon these foundations, many targets such as molecule structures\cite{ingrahamLEARNINGPROTEINSTRUCTURE2019,wangDifferentiableMolecularSimulations2020,navarroTopDownMachineLearning2023}, radial distribution function (RDF)\cite{wangLearningPairPotentials2023,thalerLearningNeuralNetwork2021,wangDMFFOpenSourceAutomatic2023,matinMachineLearningPotentials2024}, and free energy\cite{wangDMFFOpenSourceAutomatic2023} have already been used in top-down training. However, the utilization of dynamical targets such as transport coefficients and spectroscopic data have not been investigated yet.

The essential reason why dynamical data has not been used in top-down fitting roots in the fundamental obstacles of trajectory differentiation. Previously, it was believed that trajectory differentiation suffers from severe memory overflow issue as the memory cost of backpropagation scales linearly with respect computational depth (i.e., the number of MD steps in our case). Furthermore, direct trajectory differentiation leads to exploding gradients due to the chaotic character of the long time molecular dynamics\cite{ingrahamLEARNINGPROTEINSTRUCTURE2019,metzGradientsAreNot2022}.
When dealing with thermodynamic properties (i.e., density, RDF, free energy etc.), we can circumvent these issues by employing the reweighting scheme\cite{thalerLearningNeuralNetwork2021, wangDMFFOpenSourceAutomatic2023}, as thermodynamic properties are only related to the stationary distributions instead of the sampling trajectories. But for dynamical properties (i.e., diffusion coefficient, electrical conductivity, IR spectrum etc.), a deep differentiation chain through the entire trajectory seems to be inevitable. There are some attempts to the alleviate these two problems. The adjoint method inspired by the NeurODE work\cite{chenNeuralOrdinaryDifferential2019} can reduce the memory cost to constant. And some methods, such as damped backpropagation\cite{ingrahamLEARNINGPROTEINSTRUCTURE2019}, truncated backpropagation\cite{tallecUnbiasingTruncatedBackpropagation2017,aicherAdaptivelyTruncatingBackpropagation}, partial backpropagation\cite{sipkaDifferentiableSimulationsEnhanced2023}, or even the black box gradient\cite{metzUnderstandingCorrectingPathologies,parmasPIPPSFlexibleModelBased}, have been applied in simple cases in other areas to address the exploding gradient issue. However, the effectiveness of these approaches in the differentiation of dynamical properties in high-dimensional bulk MD have not been examined.

In this work,  we propose to fill this gap by combining both differentiable multistate Bennett acceptance ratio estimator (MBAR)\cite{shirtsStatisticallyOptimalAnalysis2008a} and trajectory differentiation techniques. We show that the adjoint technique\cite{pontryaginMathematicalTheoryOptimal2018,chenNeuralOrdinaryDifferential2019} is highly effective in the reversible NVE simulations typically employed in dynamical property calculation. And we show that the gradient explosion issue can be avoided by proper damping or truncation schemes, due to the reason that although single trajectory is chaotic, the evaluation of ensemble-averaged dynamical properties must be stable. Therefore, the differentiation of dynamical properties may not be as ill-conditioned as it was once believed. Eventually, using water as a simple example, we show that by combining both thermodynamic and spectroscopic data, we can boost the DFT-based ML potentials towards a significantly higher accuracy, eventually leading to more robust predictions to other properties such as RDF, diffusion coefficient, and dielectric constant. Through these test cases, we show how the details of microscopic intermolecular interactions can be learned from comprehensive experimental observations.

\section*{Results}

\subsection*{Differentiation of Dynamical Properties}

Important dynamical properties in chemistry and materials science, including both transport coefficients and spectroscopy, are always related to ensemble averaged time correlation functions (TCFs)\cite{mcquarrie76a}. For example, under the Green-Kubo formula\cite{greenMarkoffRandomProcesses1954a,kuboStatisticalMechanicalTheoryIrreversible1957}, transport coefficients $\langle O\rangle$ can be expressed as:

\begin{equation}
\langle O\rangle \propto \int_0^\infty C_{AB}(t) dt = \int_0^\infty\langle A(0)\cdot B(t)\rangle dt\label{eq:tcf}
\end{equation}

where $C_{AB}(t) = \langle A(0)\cdot B(t)\rangle$ is the TCF of observables $A$ and $B$. For example, for diffusion coefficient, both $A$ and $B$ are particle velocities, and for electrical conductivity, both $A$ and $B$ are ionic current. Meanwhile, vibrational spectroscopy is also related to TCFs via Fourier transform:

\begin{equation}
I(\omega) \propto \int_{-\infty}^\infty C_{AB}(t) e^{-i\omega t} dt\label{eq:FTIR}
\end{equation}

\begin{figure*}[htbp]
 \centering
 \includegraphics[width=0.8\textwidth]{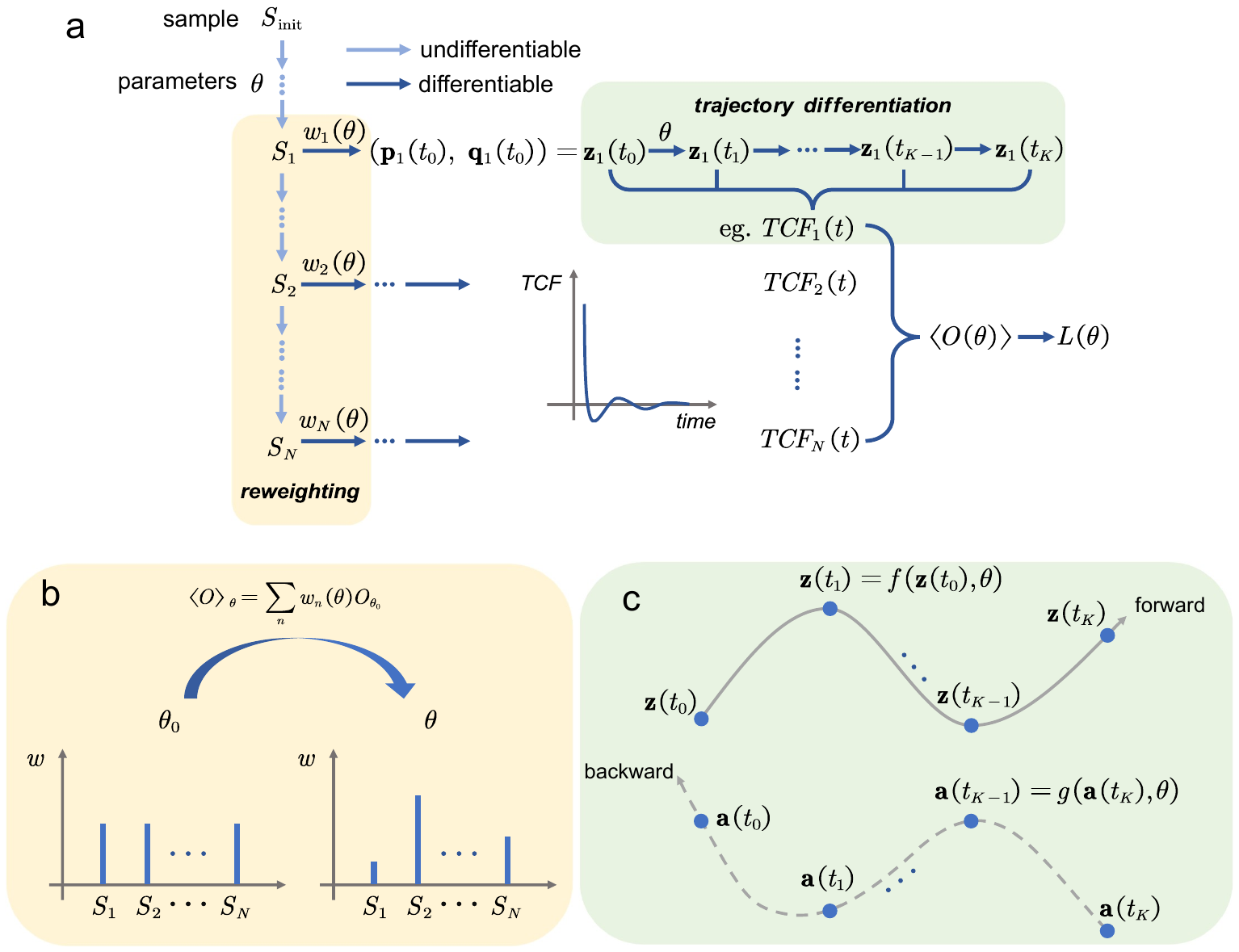}
 \caption{Differentiable computation flow for dynamical properties. (\textbf{a}) Overall work flow from initial sampling to TCF and loss function calculations. The initial sampling part does not need to be differentiable. TCF for each trajectory is generated by moving averaging over the trajectory from $\textbf{z}(t_0)$ to $\textbf{z}(t_K)$. The averaged TCF is then used to compute the target property and the loss function. (\textbf{b}) demonstrates the differentiable reweighting technique that is used to compute the thermodynamic ensemble average. (\textbf{c}) shows the how trajectory differentiation is conducted, by forward propagating state $\textbf{z}(t_i)$, followed by a backward propagation of adjoint state $\textbf{a}(t_i)$. $f$ and $g$ are their respective equations of motion, defined by Eq. \ref{eq:dyn_fwd} and \ref{eq:adj_dyn}.}
 \label{fig:scheme}
\end{figure*}

Different to thermodynamic properties, $C(t)$ is not only a function of the ensemble distribution but also related to the dynamics of trajectories. In the classical limit, the typical workflow to compute $C(t)$ is to first draw an ensemble of samples from an equilibrated NVT or NPT simulation (denoted as: $S_1,S_2,\cdots,S_N$). Then, starting from these initial samples, the time evolution of the system state (i.e., the position and momentum of each particle) can be propagated using classical NVE integrator, generating the full trajectory information: $\textbf{z}_n(t)=\left(\textbf{p}_n(t),\textbf{q}_n(t)\right)$ (here, $\textbf{p}$ denotes momenta, $\textbf{q}$ denotes positions, and $n=1,2,\dots,N$ is the initial sample index). The ensemble of trajectories is then used to compute $C(t)$, the dynamical properties, and the corresponding loss function $L$ to be optimized (see Fig. \ref{fig:scheme}a).

In this computing process, the potential parameter $\theta$ enters in both the sampling of initial states and the propagation of each trajectory, leading to two gradient terms. Taking the computation of transport coefficients as an example, when using squared deviations as loss function, it can be written as:

\begin{equation}
L=(\langle O\rangle-O^*)^2 \label{eq:loss}
\end{equation}

with:

\begin{equation}
\label{eqn:}
\begin{cases}
&\left<O\right> = \sum_{n=1}^N {w_n(\theta) O_n} \\
&O_n = \int_0^\infty A(\mathbf{z}_n(0))B(\mathbf{z}_n(t, \theta)) dt 
\end{cases}
\end{equation}

Then its gradient with respect to $\theta$ is:

\begin{equation}
\frac{\partial L}{\partial \theta}=2(\langle O\rangle-O^*)\left(\sum_{n=1}^N\frac{\partial w_n}{\partial\theta}O_n+\sum_{n=1}^N w_n\frac{\partial O_n}{\partial \theta}\right) \label{eq:grad_o}
\end{equation} 

In here, $w_n$ is the statistical weight associated with each sample $S_n$, which should be $\frac{1}{N}$ in a normal MD sampling. However, in the context of reweighting scheme\cite{shirtsStatisticallyOptimalAnalysis2008a}, $w_n$ can be viewed as a function of $\theta$, thus giving the first term in Eq. \ref{eq:grad_o}. In this work, we call this term "thermodynamic gradient", which can be evaluated easily using the differentiable MBAR technique we developed in our previous work\cite{wangDMFFOpenSourceAutomatic2023} (Fig.~\ref{fig:scheme}b). The second term, which we call "dynamical gradient", is related to the differentiation of trajectory (i.e., $\left.\frac{\partial \mathbf{z}_n(t)}{\partial \theta}\right|_{\mathbf{z}_n(0)}$) and more difficult to compute. As mentioned above, the memory cost of the naive backpropagation algorithm scales linearly with respect to computation depth (i.e. number of time steps $K$). Thus for an accurate dynamics with small time step $dt$, naive backpropagation differentiation can easily consume all the memory. In this case, the adjoint technique can be introduced as a perfect solution, reducing the memory cost of the trajectory differentiation to constant (Fig.~\ref{fig:scheme}c). 

Consider a typical NVE simulation setup, we are effectively propagating a deterministic ordinary differential equation (ODE) using the following dynamics:

\begin{equation}
\textbf{z}(t_i)=f(\textbf{z}(t_{i-1}),\theta)\label{eq:dyn_fwd}
\end{equation}

where $f$ is a NVE integrator such as Velocity-Verlet or leap-frog, both of which is rigorously time reversible: $(-\textbf{p}(t_{i-1}),\textbf{q}(t_{i-1}))=f((-\textbf{p}(t_i),\textbf{q}(t_i)),\theta)$. The adjoint variable $\mathbf{a}(t_i)$ is then defined as the differentiation of the loss function $L$ with respect to $\mathbf{z}(t_i)$, while keeping the trajectory before $t_i$ unchanged but allowing the trajectory after $t_i$ to vary accordingly:

\begin{equation}\label{eq:adj_def}
\mathbf{a}(t_i) = \left.\frac{\partial L}{\partial \mathbf{z}(t_i)}\right|_{\mathbf{z}(t_j<t_i), \theta}
\end{equation}

When $L$ is an explicit function of the entire trajectory: $L=L(\mathbf{z}(t_0), \mathbf{z}(t_1), \dots, \mathbf{z}(t_K))$, the last adjoint $\mathbf{a}(t_K)$ can be trivially computed as a direct derivation $\frac{d L}{d \mathbf{z}(t_K)}$. Starting from this point, we can then solve the reverse dynamics of $\mathbf{a}$ using chain rule:

\begin{equation}
\textbf{a}(t_{i-1})=\textbf{a}(t_i)\frac{\partial f(\textbf{z}(t_{i-1}, \theta))}{\partial \textbf{z}(t_{i-1})}+\frac{d L}{d \textbf{z}(t_{i-1})} =g(\mathbf{a}(t_i), \mathbf{z}(t_i), \theta)\label{eq:adj_dyn}
\end{equation}

Thanks to the rigorous time reversibility of $f$, the variable $\frac{\partial f(\mathbf{z}(t_{i-1},\theta))}{\partial \mathbf{z}(t_{i-1})}$ can be computed using only $\mathbf{z}(t_{i})$, without any saved information of $\mathbf{z}(t_{i-1})$. Therefore, following the dynamics given by Eq. \ref{eq:adj_dyn}, the whole adjoint trajectory can be obtained via a single backward pass of the existing trajectory. And according to previous work\cite{chenNeuralOrdinaryDifferential2019}, the total parameter gradient can be then computed along the backward pass as:

\begin{equation}
\frac{dL}{d\theta}=\sum_{i=K}^1\textbf{a}(t_i)\frac{\partial f(\textbf{z}(t_{i-1}),\theta)}{\partial\theta} \label{eq:odegrad_param}
\end{equation}

This computation process is highly efficient: the computational cost of the entire adjoint backward pass is similar to another normal MD run, with a constant memory cost that does not depend on number of steps. 

We note that this adjoint technique has been well established  to study a variety of dynamical systems\cite{lettermannTutorialBeginnerGuide2024}. However, in the field of high-dimensional MD, researchers have primarily focused on thermodynamic properties. For thermodynamic calculations, long sampling MD trajectory with heat bath is usually required. The frequently used stochastic heat baths (such as Langevin thermostat) break the time reversibility of the MD step, thus cumbering the adjoint propagation. Therefore, most previous studies\cite{wangDifferentiableMolecularSimulations2020,wangLearningPairPotentials2023} used the reversible extended Lagrangian NVT thermostat. However, the differentiation of long sampling trajectory is sub-optimal, as the reweighting scheme is much more efficient for thermodynamics. Meanwhile, in the case of dynamical property (i.e., TCF) calculations, differentiating trajectory becomes inevitable, which is a perfect application scenario for the adjoint technique. Generally, the optimization of any time reversible dynamics, such as quantum ring-polymer MD (RPMD)\cite{craig2004quantum} or LSC-IVR (linearized semiclassical initial value prepresentation)\cite{wang1998semiclassical,sun1998semiclassical} simulations, can also be conducted in a similar way. However, in this work, we only focus on the unperturbed classical dynamics, leaving the reversible quantum dynamics to our future work.

While the adjoint technique is highly efficient, the differentiation of a single trajectory is ill-conditioned in the long-time limit. The chaotic nature of the MD system leads to an exponentially increasing statistical noise with respect to time in the TCF gradient evaluations (\textit{vide infra}). To alleviate this problem, starting from the initial samples, we propagate the system trajectories in both forward and backward directions. Through this trick, we effectively double the trajectory lengths while keeping the statistical noise of gradient staying in the same magnitude. Furthermore, from the physical perspective, the time scale of chaotic explosion (e.g., the Lyapunov time) of a MD system is inherently related to how fast the system loss the information of its initial condition, which should be related to the correlation time scale that characterizes the decay rate of TCF. This understanding implies that while the statistical noise of long-time gradient inevitably explodes, the actual importance of its averaged value probably decays in a comparable time scale. Therefore, the ill-conditioned long-time differentiation can be avoided by simple truncation or damping schemes. Although a rigorous analysis is still needed to compare these two time scales, through numerical tests, we show that the truncation/damping scheme is indeed applicable in many important situations.

In this work, all calculations, including the evaluations of both thermodynamic and dynamical gradients were conducted using the DMFF package\cite{wangDMFFOpenSourceAutomatic2023}. The adjoint method was implemented in DMFF based on the leap-frog integrator, using JAX\cite{bradburyJAXComposableTransformations2018} framework.

\subsection*{Transport Coefficients}

Before digging into spectroscopy, we first test our method in the fitting of transport coefficients, which is an important category of target properties in materials design. In this work, we use the self-diffusion coefficient of Lennard-Jones (LJ) particles as a toy system to showcase our strategy. 

The potential function of a single-component LJ liquid is defined as:

\begin{equation}
V_{LJ}=\sum_{a<b} {4\epsilon\left[\left(\frac{\sigma}{r_{ab}}\right)^{12}-\left(\frac{\sigma}{r_{ab}}\right)^{6}\right]} \label{eq:lj_pot}
\end{equation}

where $\sigma$ and $\epsilon$ are the parameters to be optimized, and $r_{ab}$ is the distance between two particles $a$ and $b$. The diffusion coefficient is calculated using the Green-Kubo formula:

\begin{equation}
D=\frac{1}{3}\int_0^\infty \langle v(0)\cdot v(t)\rangle dt\label{eq8}
\end{equation}

where $\langle v(0)\cdot v(t)\rangle$is the TCF of velocity. In this test case, the diffusion coefficient computed using parameters $\sigma^*=0.34~\text{nm}$ and $\epsilon^*=0.993795~\text{kJ/mol}$\cite{rahmanCorrelationsMotionAtoms1964} is used as the fitting target. The optimization starts from an initial point of $\sigma=0.338~\text{nm}$ and $\epsilon=0.893795~\text{kJ/mol}$. The goal is to examine whether or not the gradient of diffusion coefficient can guide the optimizer to find the original point $(\sigma^*, \epsilon^*)$ in the 2D parameter space.

\begin{figure*}[htbp]
\centering
\includegraphics[width=0.7\linewidth]{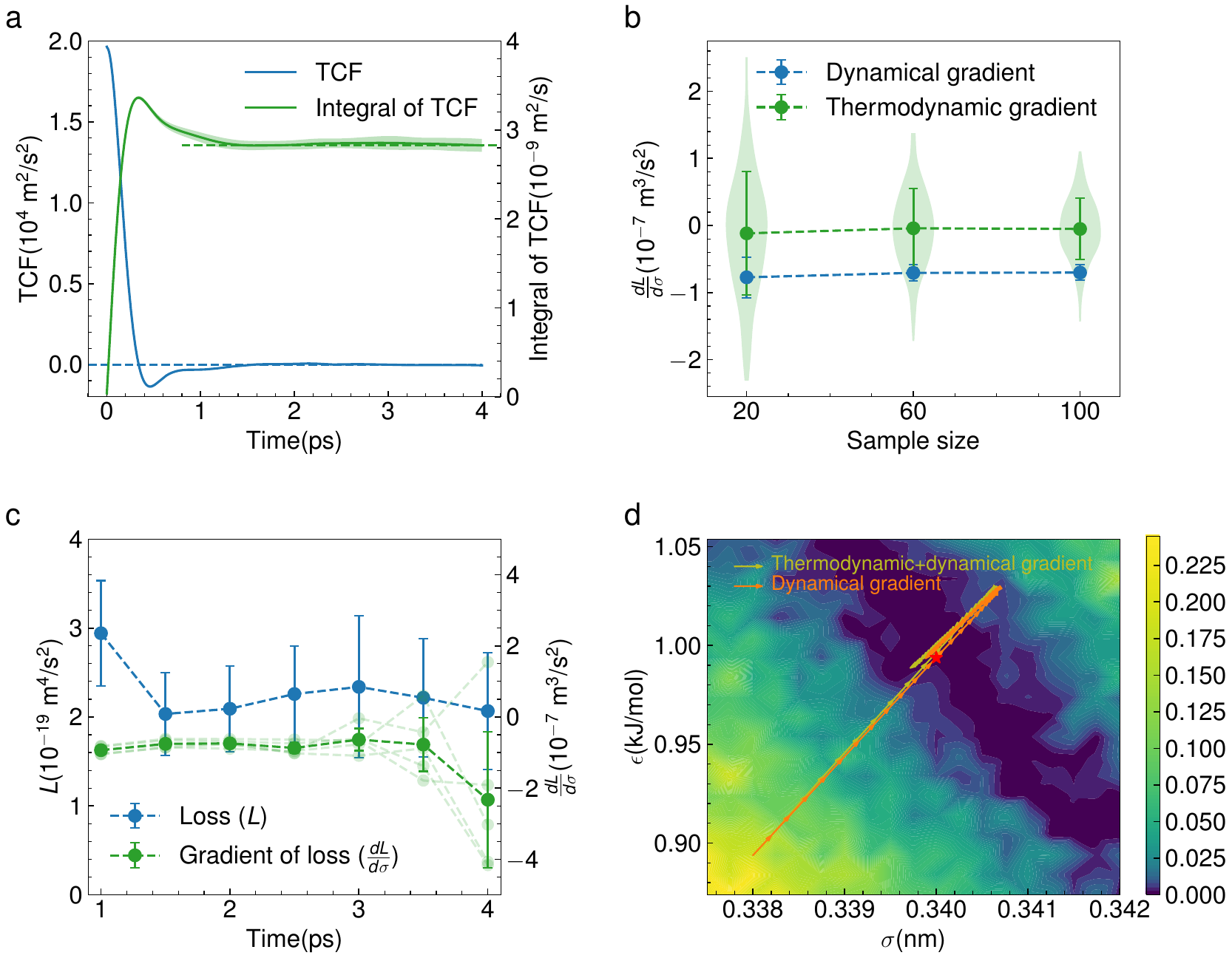}
\caption{Fitting diffusion coefficient for LJ liquid. (\textbf{a}) The convergence of the TCF and the integral of TCF over time, 5 independent trials were conducted, with the uncertainly marked by the shaded area. (\textbf{b}) Comparison between the thermodynamic and dynamical gradients as a function of sample sizes, 5 independent trials are conducted for dynamical gradient and 100 independent trials are conducted for thermodynamic gradient, giving the statistical uncertainty of their distributions.  (\textbf{c}) The convergence of loss function and its gradient, as a function of integration time. 5 independent trials are conducted to evaluate the statistical uncertainty. (\textbf{d}) The optimization paths in the parameter space, using either total gradient or the dynamical gradient only.}
\label{fig:lj_test}
\end{figure*}

Since the initial sampling process does not need to be differentiable, we use OpenMM\cite{eastmanOpenMMRapidDevelopment2017} to run the NVT simulation and randomly select 100 frames as the initial structures for the next-stage NVE runs. The integral of TCF converges within 1.5~ps with a relative error less than 2\% (Fig.~\ref{fig:lj_test}a). Therefore, to ensure convergence, we propagate each trajectory for 2~ps.

To demonstrate the stability of the gradient, we investigate the effects of the sample size (Fig.~\ref{fig:lj_test}b) and the time length (Fig.~\ref{fig:lj_test}c). First, we compare the thermodynamic and dynamical gradients by presenting their respective means and standard deviations in Fig.~\ref{fig:lj_test}b, as a function of sample sizes. As the sample size increases from 20 to 100, dynamical gradient changes from -0.77$\pm$0.30, -0.71$\pm$0.12, to -0.70$\pm$0.11; while thermodynamic gradient changes from -0.12$\pm$0.92, -0.04$\pm$0.59, to -0.05$\pm$0.46 in 10$^{-7}$ m$^3$/s$^2$. We also plot the distribution of the thermodynamic gradient using the results of 100 independent trials, as shown in the violin diagram in Fig.~\ref{fig:lj_test}b. It is evident that while the dynamical gradient is much larger and much more important in optimization, thermodynamic gradient is the major contributor to the statistical noise. This result indicates that for dynamical properties, $\theta$ primarily affects the final outcome by influencing the dynamics of trajectories, rather than the thermodynamic distribution of initial structures. And the fact that dynamical contribution dominates the total gradient indicates that we can safely neglect the thermodynamic gradient for a better numerical performance, which we will discuss in below.

Next, we compare the convergence of loss $L$ and its gradient $\frac{d L}{d \sigma}$ as functions of TCF integration time. $L$ converges within 1.5~ps, while the uncertainty of $\frac{d L}{d \sigma}$ gradually diverges after 3~ps, due to gradient explosion. Apparently, for long time dynamics, the noisy gradient is completely useless in optimization. However, as we point out above, TCF decays to zero in a faster rate, thus we can simply truncate the integration at 2~ps, so a numerically stable optimization can be conducted. 

Using the well-behaved gradient, we perform the optimization for $(\sigma,\epsilon)$ using the ADAM\cite{kingmaAdamMethodStochastic2017} optimizer. We use a learning rates of 0.0002 for $\sigma$ and 0.01 for $\epsilon$, since their gradients are in different orders of magnitude. We illustrate the contour of the loss function in the parameter space and the optimization path in Fig.~\ref{fig:lj_test}d. The location of the target parameter values are marked using the red star symbol, representing the minimum of the loss surface. Two optimization paths are shown, using either total gradient or dynamical gradient only, respectively. Throughout the entire optimization, the directions of both gradients are aligned, and both strategies successfully locate the optimal parameters within 50 loops. In fact, the dynamical gradient strategy has a better performance, as the optimizer reaches the minimum slightly faster, due to the removal of noise from the thermodynamic gradient. We find that this advantage becomes even more significant in more complicated optimizations as we will show in the next section. Nevertheless, through this simple test, we show the validity of our algorithm in differentiating dynamical properties, which then can be used to solve the inverse problem of spectroscopy. 

\subsection*{IR Spectrum}

Comparing to fitting diffusion coefficient of LJ liquid, predicting the IR spectrum of liquid water is a much more challenging task and has been extensively studied in existing literatures\cite{guillotMolecularDynamicsStudy1991,sharmaIntermolecularDynamicalCharge2005,zhangDeepNeuralNetwork2020,zhangUniversalMachineLearning2023}. Comparing to transport coefficients, IR spectrum is not only easier to obtain experimentally, but also contains much richer information about the microscopic PES. In this case, we calculate the IR spectrum using harmonic quantum correction factor (QCF):

\begin{equation}
\alpha(\omega)n(\omega)=\frac{2\pi}{3ck_BTV}\int_{-\infty}^{\infty}dt e^{-i\omega t}\langle\dot{\textbf{M}}(0)\cdot\dot{\textbf{M}}(t)\rangle\label{eq9}
\end{equation}

where $\langle\dot{\textbf{M}}(0)\cdot\dot{\textbf{M}}(t)\rangle$ is the TCF of the time derivative of total dipole, $c$ is the speed of light, $k_B$ is Boltzmann constant, $T$ is the temperature, $V$ is the volume, $\alpha(\omega)$ is the absorption coefficient, and $n(\omega)$ is the frequency-dependent refractive index. To ensure convergence, we always sample 320 independent trajectories in the following calculations. 

As the first step, we perform a simple test by fitting the IR spectrum computed using the 
q-SPC/Fw \cite{paesaniAccurateSimpleQuantum2006} force field. Similar to the LJ diffusion coefficient case, in here, we perturb the values of the force field parameters (i.e., the hydrogen charge $q$, the oxygen $\sigma$, the OH bond force constant $k_b$, and the HOH angle force constant $k_a$), and observe if IR fittings can restore their original values or not. 

As shown in Supplementary Fig. 2, the dipole derivative TCF features long echos that lasting beyond 2~ps. This long-living echo is primarily due to the intramolecular vibrations that is essentially decoupled from other motions, especially in the case of classical harmonic force fields. At first glance, such long-living mode, in combination with gradient explosion, creates tremendous difficulty in IR fitting. Indeed, in Supplementary Fig. 3, using numerical gradient as reference, we show that our scheme can only generate reliable gradient within 0.3~ps, which is certainly not enough to fully converge the IR calculation in the case of q-SPC/Fw. However, for spectroscopy fitting, a simple low-pass filter can be applied to solve this problem. We first Fourier transform the target spectrum into time domain, truncate it at 0.2~ps, then inverse Fourier transform it to obtain the real fitting target. Then we validate our fitting results using the fully-converged spectrum computed using the untruncated TCF. Employing this strategy, the optimization paths using both full gradient and dynamical gradient are shown in supplementary Fig. 4 (purterbing only $q$ and $\sigma$). In this case, the total gradient strategy fails due to the large noise from thermodynamic gradient, while the dynamical gradient strategy performs much more robust. This result once again supports the use of dynamical gradient , instead of the total gradient although it is theoretically more rigorous. When perturbing all four parameters ($q, \sigma, k_a, k_b$), the final fitting results are shown in supplementary Fig. 5 and supplementary Table 1. As shown, even though we use only the low-pass filtered spectrum as our fitting target, the fitted potential restores the parameter values and the unfiltered IR spectrum quite well. 

Physically, the q-SPC/Fw IR fitting case represents a general situation that the trajectory differentiation may fail. That is, if there exists multiple modes in one system with separating time scales, then the chaotic nature of short-living modes may hinder the differentiation of other long-living modes. The low-pass filter approach we adopt in this work is effectively using the short-time dynamics to imply the long-time behavior. This approach is shown to perform reasonably well in our case, giving that both the short- and the long-time dynamics are controlled by the same PES, and can both benefit from its improvement.    

Giving the success in q-SPC/Fw fitting, we next switch to train a ML potential using experimental IR data\cite{bertieInfraredIntensitiesLiquids}. To directly compare with experiment, we also need an accurate dipole surface besides a PES. In this work, we use the Deep Dipole\cite{zhangDeepNeuralNetwork2020} model trained at PBE0 level of theory to provide the dipole surface. For simplicity, we keep the dipole model fixed in the following study, assuming its accuracy. We note that technically, the dipole model can be optimized together with the PES model, but a separate fitting with designated dipole labels is preferred. We start our test by pretraining an embedded atom neural network (EANN) potential\cite{zhangEmbeddedAtomNeural2019} using the dataset generated in previous work\cite{koIsotopeEffectsLiquid2019}. This dataset contains the energies and forces data of 64 water molecules in periodic boundary conditions, computed at dispersion corrected PBE0-TS (Tkatchenko-Schefler) level of theory. We then refine this bottom-up machine learning model using either density data alone, or a combined target of density and IR data.

The combined density and IR loss function is defined as:

\begin{equation}
L=w_{\rho}(\rho-\rho^*)^2+\sum_{i} w_I(\omega_i)[I(\omega_i)-I^*(\omega_i)]^2\label{eq10}
\end{equation}

where $\rho$ is the density, $I(\omega_i)=\alpha(\omega_i)n(\omega_i)$ is the spectrum, and values marked with asterisk (*) are experimental references. The weights $w_{\rho}$ and $w_I(\omega_i)$ are used to balance the contributions of density and spectrum respectively. $w_{\rho}$ is set to be $10^3~\text{cm}^6/\text{g}^2 $, and $w_I$ is set to be frequency-dependent. It is noted that the IR spectrum in the high frequency region (i.e., $>1000~\text{cm}^{-1}$) shows significant nuclear quantum effects (NQEs), which is not supposed to be accurately described by classical dynamics. Furthermore, the strong peak at 3400~cm$^{-1}$ is likely to dominate the loss function if a uniform weight scheme is utilized. Therefore, we put particular emphasize on the low-frequency region ($<1000~\text{cm}^{-1}$), and increase $w_I(\omega)$ of the low-frequency region by a factor of 10, setting it to be $10^{-8}~\text{cm}^2$, in comparison with $10^{-9}~\text{cm}^2$ in the high-frequency region.  For optimization, we utilize the ADAM optimizer with a learning rate of 0.00005  for all the parameters. The EANN potential contains thousands of parameters, with a much higher risk of overfitting compared to classical potential. So we keep monitoring the change of self-diffusion coefficient in the entire optimization process, using it as the validation data to determine the ending point of the optimization cycle.

\begin{figure*}[htbp]
\centering
\includegraphics[width=0.7\linewidth]{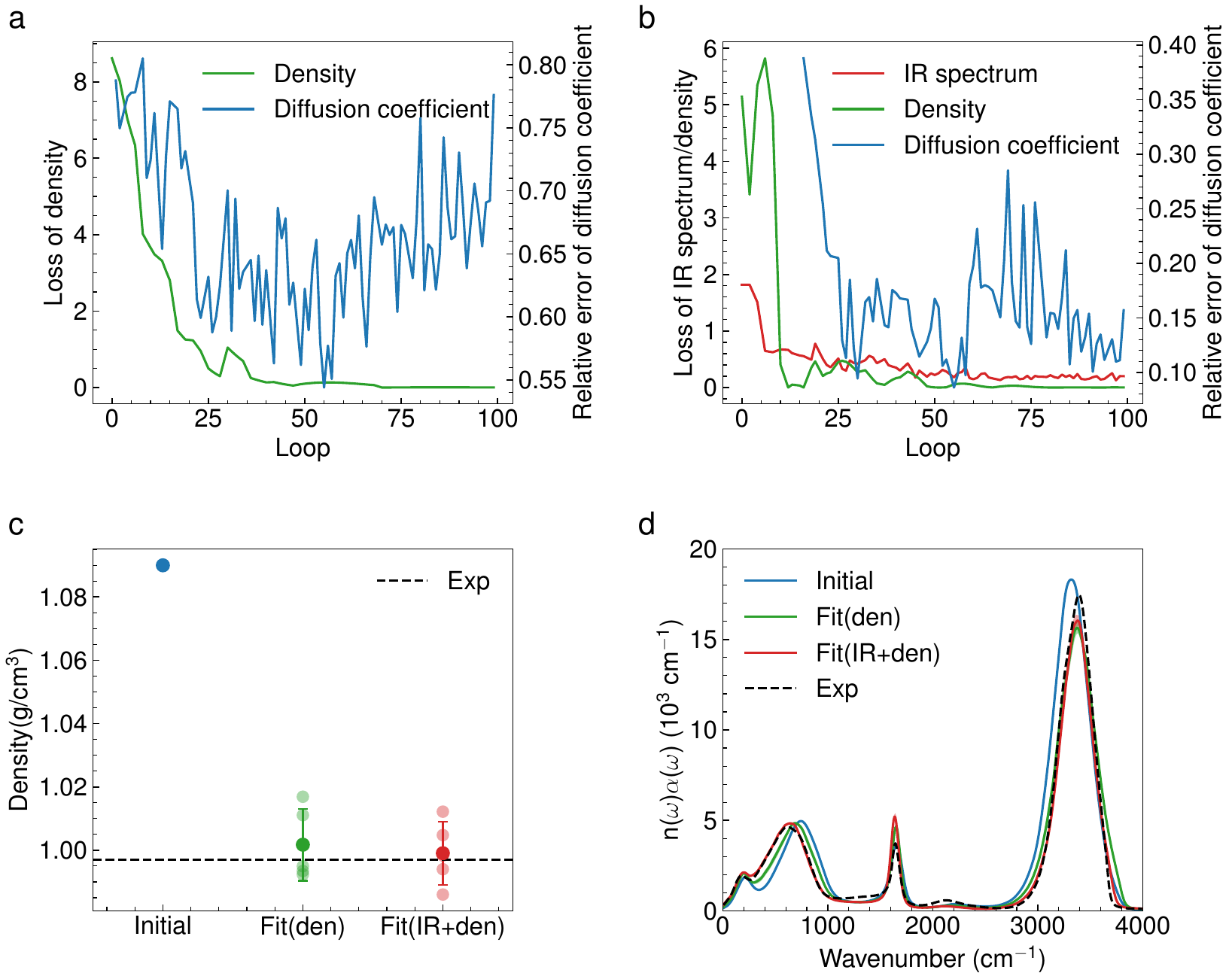}
\caption{Fitting IR spectrum and density for liquid water. (\textbf{a}) Optimization process using only density as target; (\textbf{b}) Optimization process using both IR spectrum and density targets. The change of diffusion coefficient error is also plotted along the side of density and IR spectrum errors. (\textbf{c}) Predicted densities before and after the optimization. Results from five independent trials are shown, comparing with the experimental value marked by dashed line. (\textbf{d}) IR spectrum before and after optimization, comparing with experimental results\cite{bertieInfraredIntensitiesLiquids}. Five independent optimizations were conducted to compute the uncertainty of the final results, which are represented by the shaded area.}
\label{fig:ir_opt}
\end{figure*}

The optimization process using combined density and IR target is depicted in Fig.~\ref{fig:ir_opt}b, showing gradually decreasing losses in both IR spectrum and density. The initial PBE0-TS level EANN model features large error in density prediction. Therefore, the density loss dominates the optimization and reaches convergence within the first 10 optimization loops. After that, the IR loss becomes the main driving force for optimization, where some competitions between density and IR losses can be observed. As the independent validation, the error of diffusion coefficient keeps a general decreasing trend up to the 50'th loop, before starting to oscillate, indicating the risk of overfitting. Therefore, as a general practice, we always pick the parameters giving the smallest error of self-diffusion coefficient within the first 100 loops as our final model. For example, in the case shown in Fig.~\ref{fig:ir_opt}, the parameter from loop 58 is selected for further tests.  In supplementary Fig. 6, we present the evolution of the IR spectrum versus the experimental reference  every 5 loops, showing the progress of the optimization.

To evaluate the stability of the procedure, we conduct 5 independent trial optimizations, the final performances of which are shown in Fig.~\ref{fig:ir_opt}cd. It is shown that better performances are achieved in both density and IR spectrum when combined density and IR target is used, in comparison with the pure density fitting. This demonstrates the effectiveness of dynamical information in the top-down fine-tuning. Even more significant effects can be observed when we transfer the refined potential to predict other properties, as we will show in below.

\subsection*{Transferability to Other Properties}

\begin{figure*}[htbp]
\centering
\includegraphics[width=0.7\linewidth]{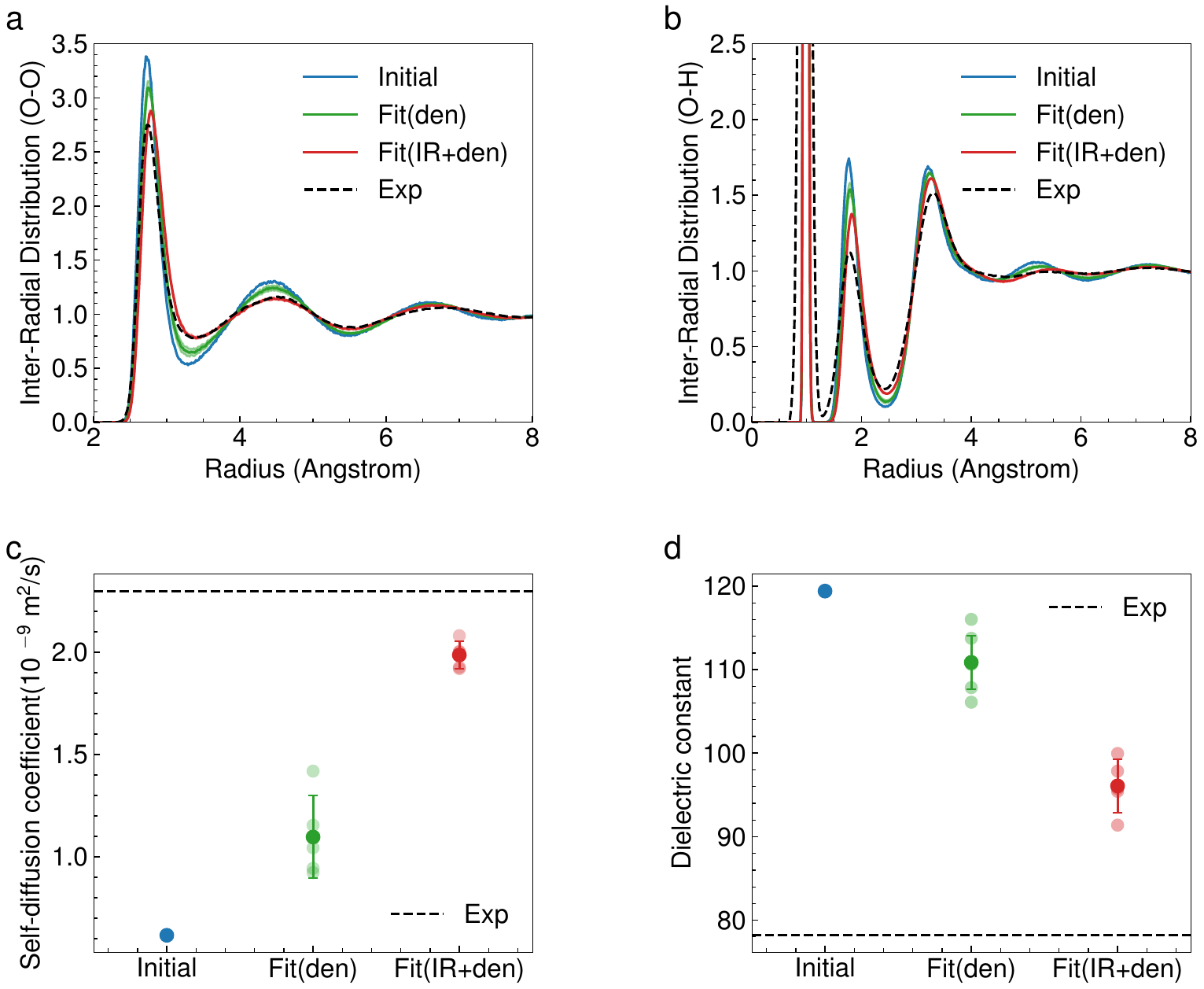}
\caption{Performances of refined ML PES  for liquid water, in comparison with the initial bottom-up model and experimental values:  (\textbf{a}) O-O and (\textbf{b}) O-H RDFs comparing with experimental data (Ref.\cite{soperRadialDistributionFunctions2000}). (\textbf{c}) Self-diffusion coefficients in comparison with with experimental values (Ref.\cite{holzTemperaturedependentSelfdiffusionCoefficients2000}). (\textbf{d}) Dielectric constants in comparison with experimental values (Ref.\cite{malmbergDielectricConstantWater1956}), 5 independent optimizations were conducted, the results of which are shown using shadow points or areas.}
\label{fig:benchmark}
\end{figure*}

To demonstrate that the optimization is not only a numerical fitting but also learns real physics from the spectrum, we use the optimized PES to compute other properties that are not included in the fitting target. The RDF, self-diffusion coefficient, and dielectric constant are used, giving a comprehensive evaluation to the model's performance on both the thermodynamic and the dynamical behaviors of bulk liquid water. As shown in Fig.~\ref{fig:benchmark}ab, the potentials refined with density and IR spectrum perfectly reproduce the experimental O-O and O-H RDFs\cite{soperRadialDistributionFunctions2000}, performing significantly better than the potential fitted to density only. Combined density and IR fitting yields a self-diffusion coefficient of $1.99\pm0.07\times10^{-9}\text{m}^2/\text{s}$, which is comparable to the experimental value of  $2.299\times10^{-9}\text{m}^2/\text{s}$ at 298 K\cite{holzTemperaturedependentSelfdiffusionCoefficients2000}, and is much more accurate than the pure density fitting results (Fig.~\ref{fig:benchmark}c). Moreover, the potentials refined by combined density and IR targets give a dielectric constant prediction of 96.1$\pm$3.2, which, although being higher than the experimental value of 78.3\cite{malmbergDielectricConstantWater1956}, still represents a significant improvement over the initial bottom-up potential (119.4) and the potential refined using only density (110.9$\pm$3.2). The residual error in dielectric constant may be attributed to the stopping criterion we used in the optimization process. Since we are using diffusion coefficient error to pick the final model, the resulting potential is perhaps biased towards a better description to diffusion. In general, more types of fitting targets should lead to more robust optimization results. However, the optimal set of targets for an ideal top-down PES refinement process is subject to our future research. Nevertheless, the results in Fig. \ref{fig:benchmark} clearly shows the essential impacts of spectroscopic data, which, further considering its experimental accessibility, should be a critical component in the future fine-tuning workflow.

\begin{figure*}[htbp]
\centering
\includegraphics[width=0.8\linewidth]{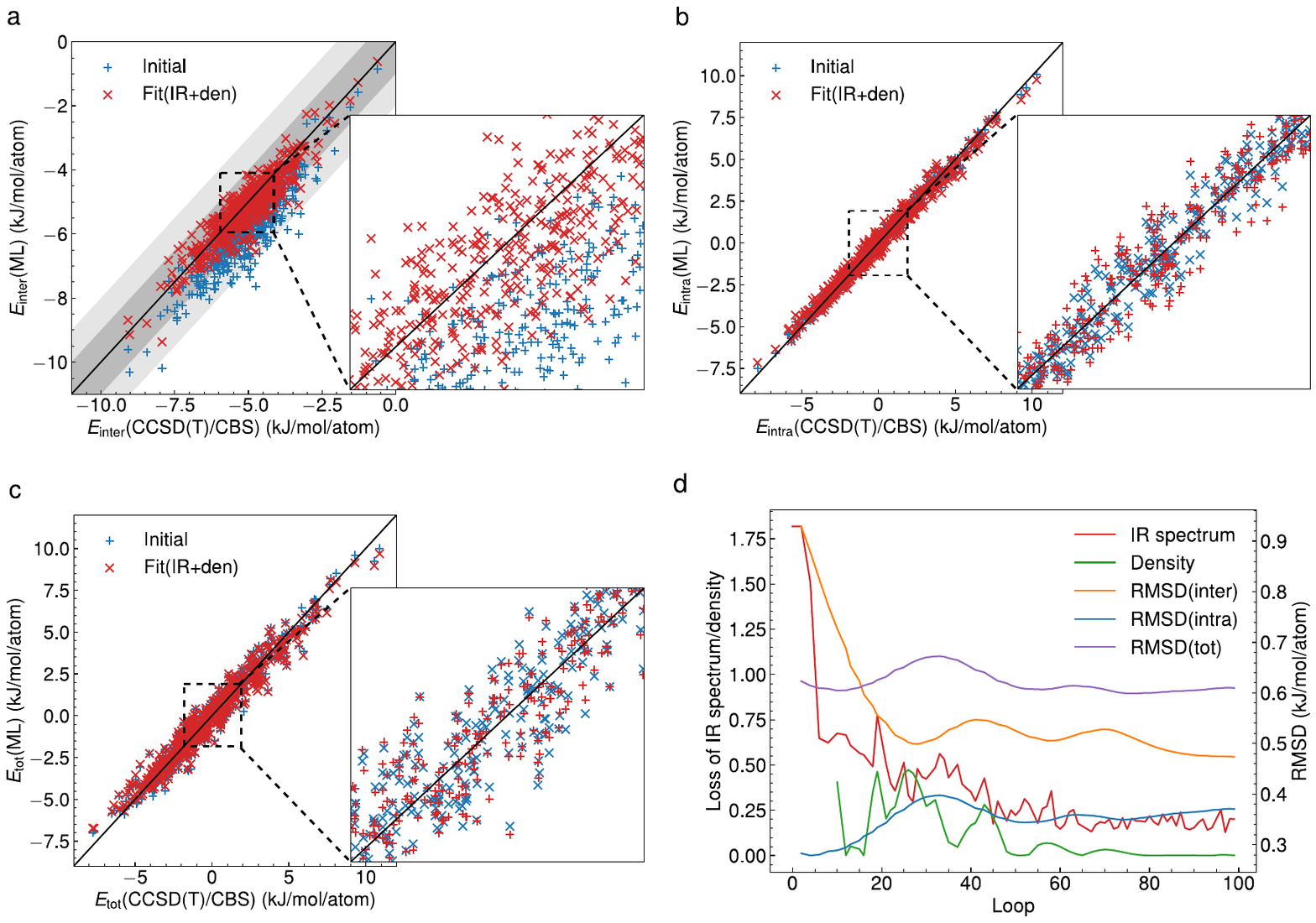}
\caption{Performance of ML PES on the CCSD(T)/CBS dataset (500 octamers) in predicting the \textbf{a} intermolecular energy, \textbf{b} intramolecular energy and \textbf{c} total energy. \textbf{d} shows the RMSD changes of intermolecular, intramolecular and total energy in optimization}
\label{fig:ab_initio}
\end{figure*}

To further verify and understand the physical effects of the top-down fine-tuning process from the microscopic perspective, we compare the refined potential with the CCSD(T)/complete basis set (CBS) calculation results (i.e., the "golden standard" in computational chemistry). The test set is taken from our previous work\cite{yangTransferrableRangeseparatedForce2022}, containing the energies of 500 octamer and 4000 pentamer water clusters. The total energy is decomposed into inter- and intra-molecular contributions, which are compared separately in Fig. \ref{fig:ab_initio}ab. The corresponding root mean squared deviations (RMSD) are also given in Table \ref{tab:rmsd_abinitio}. The RMSD of the total energy does not show any improvement after the refinement using IR. However, a decomposition of energy shows that the combined density and IR refinement significantly decreases the RMSD of inter-molecular energy, automatically correcting a large systematic error presented in the bottom-up model (Fig. \ref{fig:ab_initio}a). Meanwhile, the performance of the refined potential in intra-molecular energy is worse, counterbalancing the improvement in the inter-molecular term. In Fig. \ref{fig:ab_initio}d, we show how the energy RMSDs change during the optimization process, plotted along the side of the change of both density and IR losses. One can clearly see the deterioration of intra-molecular energy and the improvement of inter-molecular energy, both primarily driven by the optimization of IR spectrum instead of density. This discrepancy can be attributed to NQE. The low frequency band, which is related to molecule translations and librations, is primarily controlled by intermolecular interactions, while the high frequency bands in 1500 cm$^{-1}$ and 3400 cm$^{-1}$ are dominated by intramolecular vibrations. Therefore, NQE, which mainly affects the high frequency motions in room temperature, has a larger impact to the refinement of intramolecular potentials. Conceptually, it can be viewed that the IR refinement process is actually learning the potential of mean force (PMF) of adiabatic centroid MD (CMD), instead of the real physical potential. Luckily, in principle, this PMF is exactly what is needed when predicting experimental properties using classical MD. In supplementary Fig. 7, we also show the same results using only density fitting target, which is significantly inferior to the results of combined density and IR fitting. Once again, this observation highlights the importance of spectroscopic data, which brings in new physical information in the fine-tuning process.

\begin{table*}[htbp]
\centering
\caption{\label{tab:rmsd_abinitio}RMSD (in kJ/mol/atom) of ML potential on CCSD(T)/CBS dataset (standard deviation is shown in parentheses)}
\begin{tabular}{cccccccccc}
\hline
               & \multicolumn{3}{c}{Inter}        & \multicolumn{3}{c}{Intra}        & \multicolumn{3}{c}{Tot}          \\ \hline
               & Initial & Fit(den) & Fit(IR+den) & Initial & Fit(den) & Fit(IR+den) & Initial & Fit(den) & Fit(IR+den) \\ \hline
Octamers& 0.97    & 0.71(4)  & 0.49(2)     & 0.29    & 0.31(2)  & 0.36(3)     & 0.64    & 0.62(1)  & 0.62(2)     \\
Pentamers& 0.95    & 0.78(2)  & 0.67(1)     & 0.38    & 0.41(2)  & 0.47(4)     & 0.91    & 0.91(2)  & 0.92(3)     \\ \hline
\end{tabular}
\end{table*}

\section*{Discussion}

In this work, we address the inverse problem on how to infer microscopic molecular interactions from experimental dynamical data, especially the spectroscopic data. We tackle this problem using the automatic differentiable technique, which is conventionally believed to be impractical due to the memory cost and the gradient explosion issues. Through the application of adjoint method and appropriate gradient truncation strategies, we show that both the memory and gradient explosion issues can be alleviated. It is further shown that using only dynamical gradient, instead of total gradient, we can significantly reduce the statistical noise in the differentiation of dynamical properties, thus stabilizing the optimization process. Combining all these techniques, both transport coefficients and spectroscopic data can be differentiated efficiently, thus enabling a highly automatic and scalable PES optimization workflow based on dynamical properties. Using liquid water as an example, we show that a DFT-based ML potential can be promoted towards the CCSD(T)/CBS accuracy by learning from experimental spectroscopic data. And the top-down refined potential generates much more accurate predictions to other thermodynamic and dynamical properties, compared to its bottom-up precursor fitted at DFT level of theory. This work thus presents a general approach to improve the accuracy of DFT-based ML potential. More importantly, it provides an extremely efficient tool to extract and exploit the detailed microscopic information from spectroscopy, which was inaccessible via human efforts. 

Through our work, we can now use a much wider range of properties as our fitting targets to perform top-down PES fine-tuning. However, some important questions remain unresolved, such as: what is the optimal combination of fitting targets? And what is the most economic and effective optimization workflow for ML PES development? We will explore the answers to these questions in our future study. Moreover, differentiating long time dynamics is still problematic, especially when some strongly chaotic modes interfere with other modes with long time memory. To address these issues, one should probably be more focused on a few most important degrees of freedom, rather than performing a full dimensional trajectory differentiation. Along this line, some efforts exist focusing on the stochastic Langevin dynamics\cite{darveComputingGeneralizedLangevin2009,vroylandtLikelihoodbasedNonMarkovianModels2022,xieInitioGeneralizedLangevin2024} or the Fokker-Planck equations in the reduced-dimensional space. But how to use these techniques in the inverse problem of top-down parameter fine-tuning is still an open question.   

\section*{Methods}

\textbf{Time Correlation Function (TCF)}

In classical MD, TCF ($\langle A(0)\cdot A(t)\rangle$) is usually calculated as:
\begin{equation}
\langle A(0)\cdot A(t)\rangle=\frac{1}{N}\sum_{n=1}^N[\frac{1}{N_{\tau}}\sum_{\tau}^{N_\tau} A_n(\tau)\cdot A_n(t+\tau)]\label{eq11}
\end{equation}
where $N$ is the sample size and $\frac{1}{N}\sum_{n=1}^N$ represents ensemble average, and $\frac{1}{N_{\tau}}\sum_{\tau}^{N_\tau}$ represents moving average along time. Here, $A$ is a physical quantity derived from the system state. For example, $A$  is velocity in diffusion coefficient calculations, instantaneous current in electrical conductivity calculations, and total dipole moment in IR spectrum calculations. 

In the moving average scheme, any two time points $t_i$ and $t_j$ are coupled, thus the evaluation of a single-point gradient $\frac{dL}{d\textbf{z}(t_i)}$ involves the $A(t_j)$ information of all $t_j$. Therefore, one has to save the entire $A(t)$ trajectory if the moving average scheme is employed. In the case of diffusion coefficient calculations, that means storing  the full dimensional velocity trajectory in GPU memory, which is difficult. Therefore, no moving average is used in the diffusion coefficient example. Meanwhile, in the IR case, the dimensionality of the total dipole moment derivative ($\dot{\mathbf{M}}(t)$) is much lower, so the storage of a full $\dot{\mathbf{M}}$ trajectory is feasible. In this case, moving average scheme is employed to enhance statistics.

 \textbf{Reweighting technique}
 
In the theoretical framework of reweighting, we are using samples $S_1,S_2,\cdots,S_N$ drawn from one ensemble (defined by parameters $\theta_0$) to evaluate the averages in another ensemble (defined by parameters $\theta$). Therefore, each sample carries a thermodynamic weighting factor $w_n(\theta)$, the formula of which is well known from previous studies\cite{thalerLearningNeuralNetwork2021,wangDMFFOpenSourceAutomatic2023}:
\begin{equation}
w_n(\theta)=\frac{\exp[-\beta(U_n(\theta)-U_n(\theta_0)]}{\langle \exp[-\beta(U(\theta)-U(\theta_0)]\rangle_{\theta_0}}\label{eq12}
\end{equation}
where $\beta=\frac{1}{k_BT}$, $k_B$ is Boltzmann constant, and $T$ is the temperature, $U_n(\theta)$ is the effective energy of $S_n$ computed by $\theta$, $\langle\rangle_{\theta_0}$ is the average in the sampling ensemble. Then the ensemble average of  observable $\langle O\rangle$ in $\theta$ is simply,
\begin{equation}
\langle O\rangle=\sum_i^N w_i(\theta)O_i(\theta)\label{eq13}
\end{equation}
which can be easily differentiated using AD technique.
 
 \textbf{MD simulations} 

For LJ particles, the self-diffusion coefficient was computed in a cubic box containing 346 argon atoms at 94.4 K with a density of 1.374 g/cm$^3$. In the NVT simulation, the LangevinMiddleIntegrator in OpenMM\cite{eastmanOpenMMRapidDevelopment2017} was employed, using a time step of 1 fs and a friction coefficient of 0.1 ps$^{-1}$. The cut-off radius was set to 10~\AA. Initially, the system was equilibrated for 100 ps, followed by the sampling of 100 states at 2 ps intervals during the subsequent 200 ps simulation. These states served as initial configurations for the next stage of NVE simulation. For the NVE simulations, a time step of 1 fs was utilized to run 2 ps trajectories. The trajectories were then differentiated using the DiffTraj adjoint propagation module (with the leap-frog integrator) implemented in DMFF\cite{wangDMFFOpenSourceAutomatic2023}.

For liquid water, the density, self-diffusion coefficient, and dielectric constant were all computed in a cubic box containing 216 water molecules. The IR spectrum was obtained using a box with 64 water molecules, and RDFs were computed using a box with 1018 water molecules. A timestep of 0.5 fs was used for all water simulations, with the initial density consistently set at 1.0 g/cm$^3$. All NPT simulations were performed using i-PI\cite{kapilIPIUniversalForce2019a}, interfacing with the force engines implemented in DMFF\cite{chenPhyNEONeuralNetworkEnhancedPhysicsDriven2024}. NPT simulations utilized the Bussi-Zykova-Parrinello barostat with a timescale of 100 fs and the Langevin thermostat with a timescale of 10 fs to control pressures and temperatures, respectively. All NVT simulations employed a self-implemented Langevin thermostat\cite{bussiAccurateSamplingUsing2007} with a friction coefficient of 1.0 ps$^{-1}$, and all NVE simulations were conducted using the leap-frog integrator implemented in the DiffTraj module of DMFF.

In the optimization stage, for the density, the simulation time length was set to 100 ps, with 100 samples taken from the last 50 ps of the trajectory to obtain the averaged density. For the IR spectrum, 320 samples were obtained from a 64 ps NVT simulation. Each NVE trajectory propagated in both forward and backward directions, resulting in a total trajectory length of 0.4 ps. The initial structures for density and IR spectrum calculations were obtained from the last frame of the trajectory from the previous optimization loop.

In the validation stage, for the density, a 1 ns NPT simulation was run, using the last 250 ps to determine the density. For the IR spectrum, a 320 ps NVT simulation was conducted, with 320 samples taken from the last 160 ps. Subsequently, 5 ps NVE simulations were performed to obtain the TCF of 2~ps, which was integrated to get the IR spectrum. For the self-diffusion coefficient, a 200 ps NVT simulation was executed, sampling 100 samples from the last 100 ps. Following this, 20 ps NVE simulations were conducted to obtain the TCF of 2~ps, which was integrated to determine the diffusion coefficient. For the dielectric constant, a 10 ns NVT simulation was run, using the last 9 ns to calculate the dielectric constant. For RDFs calculation, a 100 ps NVT simulation was carried out, with the last 40 ps used to compute the RDF.

\textbf{ML potential energy surface and dipole surface} 

The potential energy surface was trained using EANN model\cite{zhangEmbeddedAtomNeural2019} on a dataset containing 6812 configurations of 64 water molecules at the PBE0 level from Ref. \cite{koIsotopeEffectsLiquid2019}. We divided the dataset into training and testing sets with a ratio of 0.9:0.1. The EANN model utilized a neural network with two hidden layers, each containing 64 neurons. The initial learning rate was set to 0.001 and decayed by a factor of 0.5. Twelve radial functions and $L$ up to 2 were used to construct EAD features with a cut-off radius set to 6~\AA. The weight ratio of the loss for energy and force was set to 2:1. After training, the RMSD of the energy and force from the EANN model was 0.0668 kJ/mol/atom and 7.3548 kJ/mol/\AA, respectively, on the test dataset, as shown in Supplementary Fig. 1.

The dipole surface was trained using the Deep Dipole model\cite{zhangDeepNeuralNetwork2020} on a dataset containing 7299 configurations of 64 water molecules at the PBE0 level from Ref. \cite{zhangDeepNeuralNetwork2020}. The se\_a descriptor was used to train the Deep Dipole model. The item neurons set the size of the descriptors and fitting network to [25, 50, 100] and [100, 100, 100], respectively. The components in local environment to smoothly go to zero from 5.8 to 6~\AA. In the loss function, the pref was set to 0.0 and pref\_atomic was set to 1.0. The starting learning rate, decay steps, and decay rate were set to 0.01, 5000, and 0.95, respectively. The finial training error was below 2.0$\times10^{-3}$~\AA.

\section*{REFERENCES}

\bibliography{dynamic}

\section*{Acknowledgements}

The authors thank Dr. Lei Wang from Institute of Biophysics, Chinese Academy of Sciences, Mr. Weizhi Shao from Tsinghua University, and Dr. Xinzijian Liu from DP Technology for helpful discussions.

\section*{Author contributions statement}

K.Y. designed and conceptualized the study. B.H. implemented the method and performed all calculations. B.H. and K.Y. analyzed and interpreted results, and wrote the manuscript.

\section*{Data availability}

The result data from all the simulations in this study are provided within the paper or in the Supplementary Information file. The trained models (EANN and Deep Dipole) are available at \url{https://github.com/ice-hanbin/dynamical-fitting}, and the training data are available from the authors upon reasonable request.

\section*{Code availability}

The code for the DiffTraj module has been implemented in DMFF \cite{wangDMFFOpenSourceAutomatic2023}. The two optimization cases (transport coefficients and IR spectrum) are available at \url{https://github.com/ice-hanbin/dynamical-fitting}.

\end{document}